\documentclass[aip,jcp,showpacs,showkeys,preprint,eqsecnum,tightenlines]{revtex4-1}

\usepackage{amssymb}
\usepackage{amsmath}
\usepackage{amsfonts}
\usepackage{graphicx}
\usepackage{amsbsy}
\usepackage{color}
\usepackage{bm}
\usepackage{bbm}

\newcommand{\bq}{\begin{eqnarray}}
\newcommand{\eq}{\end{eqnarray}}
\newcommand{\bqn}{\begin{eqnarray*}}
\newcommand{\eqn}{\end{eqnarray*}}

\newcommand{\rr}{\mathbf{r}}
\newcommand{\kk}{\mathbf{k}}
\newcommand{\pp}{\mathbf{p}}
\newcommand{\qq}{\mathbf{q}}
\newcommand{\RR}{\mathbf{R}}

\begin{document}
\title{The moment sum-rules for ionic liquids at criticality}

\author{Riccardo Fantoni}
\email{rfantoni@ts.infn.it}
\affiliation{Universit\`a di Trieste, Dipartimento di Fisica, strada
  Costiera 11, 34151 Grignano (Trieste), Italy}


\date{\today}

\begin{abstract}
We discuss the first three well known moment charge-charge sum-rules
for a general ionic liquid. For the special symmetric case of the
Restricted Primitive Model, Das et al. [{\sl Phys. Rev. Lett.} {\bf
107}, 215701 (2011)] has recently discovered, through Monte Carlo
simulations, that the Stillinger-Lovett or second-moment sum-rule
fails at criticality. We critically discuss a possible explanation for
this unexpected behavior. On the other hand the fourth-moment sum-rule
turns out to be able to account for the results of the simulations
at criticality.   
\end{abstract}

\maketitle
\section{Introduction}
\label{sec:introduction}

It is well known that among all possible long-range pair-potentials,
it is only in the Coulomb case that the decay law of the correlations
faster then any inverse power is compatible with the structure of
equilibrium equations (such as the Born-Green-Yvon). Under the
exponential clustering  
hypothesis for charged fluids, a number of exact sum-rules on the
correlation functions can be obtained \cite{Martin1988}. Of particular
relevance is the Stillinger-Lovett second-moment charge-charge
sum-rule which is equivalent to the 
property that the inverse dielectric function vanishes in the limit of
small wavenumbers. When this condition holds the fluid completely
shields any external charge inhomogeneity and behaves as a conducting
medium. 

In a recent work Das, Kim, and Fisher \cite{Das2011,Das2012} found out,
through finely discretized grand canonical Monte Carlo simulations,
that in the Restricted Primitive Model (RPM) of an electrolyte, the
second- and fourth-moment charge-charge sum-rules, typical for ionic
fluids, are violated at criticality.  
For a 1:1 equisized charge-symmetric hard-sphere electrolyte their
grand canonical simulations, with a new finite-size scaling device,
confirm the Stillinger-Lovett second-moment sum-rule except, contrary
to current theory \cite{Stell1995}, for its failure at the critical
point $(T_c,\rho_c)$. Furthermore, the $k^4$ term in 
the charge-charge correlation or structure factor $S_{ZZ}(k)$
expansion is found to diverge like the compressibility when $T\to T_c$
at $\rho_c$. These findings are in evident disagreement with available
theory for {\sl charge-symmetric} models and, although their results are
qualitatively similar to behavior expected for {\sl charge-asymmetric}
systems \cite{Stell1995}, even a semiquantitative understanding
has eluded them.

Starting from the Ornstein-Zernike equation and extending at {\sl all}
densities the small density diagrammatic \cite{Hansen-McDonald}
property for the partial direct correlation functions of behaving as
$1/r$ in the $r\to \infty$ limit, it is possible to arrive quickly to
the second- and fourth-moment sum rules even if the fourth-moment one
will not be expressed in terms of just thermodynamic functions.

The second- and fourth-moment sum-rules are rigorously derived
starting from the Born-Green-Yvon equations and the exponential
clustering hypothesis by Suttorp and van Wonderen
\cite{Suttorp1987,Wonderen1987,Suttorp2008} for a thermodynamically
stable ionic mixture made of pointwise particles of charges all of the
same sign immersed in a neutralizing background, the ``Jellium''. The
same sum-rules must hold also when we allow in the ionic mixture the
presence of mobile charges of {\sl both signs}, which requires to
consider a pair-potential regularization in order to prevent opposite
charges collapse. 

In this work we critically discuss the numerical findings of Das, Kim,
and Fisher \cite{Das2011} at the light of the above mentioned
analytical work of Suttorp and van Wonderen
\cite{Suttorp1987,Wonderen1987,Suttorp2008} and of a recent result of
Santos and Piasecki \cite{Santos2015} proving the long range behavior
of the $n$-body correlation functions of a general fluid at his
gas-liquid critical point.  
   
\section{The ionic fluids model}
\label{sec:model}

A multi-component ionic mixture of an electrolyte is made of mobile
charges whose 
$\mu$ component (the particles of species $\mu$) has molar fraction
$x_\mu$ and charge $z_\mu e$, here $e$ is the unit of charge and
$z_\mu$ are integer numbers. So we may, in general, have
charges of both signs. One is generally interested in studying a
neutral state (since matter around us is neutrally charged). This can
be obtained in the event that $\sum_\mu x_\mu z_\mu=0$. Otherwise is
necessary the addition of a neutralizing uniform background of charge
density $-\rho e\sum_\mu x_\mu z_\mu$, with $\rho$ the number density
of the system of charges. A particularly simple case id the Restricted
Primitive Model (RPM) where we have only two components with
$x_1=x_2=1/2$ and $z_1=-z_2=1$ (without a background). 

The Hamiltonian of a multi-component ionic mixture consisting of $s$
components, confined in a region $\Omega\subset\mathbbm{R}^{3}$ of
volume $V$, is
\bq \label{model}
H&=&\sum_{i=1}^N\frac{p^2_{i}}{2m_{\alpha_i}}+
U(\rr_1,\ldots,\rr_N),\\ \label{modelU}
U&=&\frac{1}{2}\sum^\prime_{i,j}z_{\alpha_i}
z_{\alpha_j}v_{\alpha_i\alpha_j}(|\rr_{i}-\rr_{j}|),   
\eq 
with $p=|\pp|$ and the pair-potential
\bq \label{pp}
v_{\mu\nu}(r)=v^c(r)+v^{sr}_{\mu\nu}(r),
\eq
where $r=|\rr|$, $v^c$ is the bare Coulomb potential 
\bq \label{vc}
v^c(r)&=&\frac{e^2}{V}\sum_{\kk(\neq \mathbf{0})}
\frac{4\pi}{k^2} e^{i\kk\cdot\rr},
\eq
and $v^{sr}$ is a short-range regularization assumed integrable on
$\mathbbm{R}^3$ which includes the local repulsion effect needed to
enforce thermodynamic stability \cite{Ruelle} when we allow for the
presence of particles of opposite charge in the mixture. A first soft
regularization can be chosen as 
\bq
v^{sr}_{\mu\nu}(r)=-\frac{e^2}{r}e^{-r/d_{\mu\nu}},
\eq
where the lengths $d_{\mu\nu}$ control the exponential decay at large
distances. A second regularization amounts to introduce hard-cores,
namely 
\bq \label{modelhc}
v^{sr}_{\mu\nu}(r)=\left\{\begin{array}{ll}
\infty     & r<\sigma_{\mu\nu}\\
0          & r>\sigma_{\mu\nu}
\end{array}\right.,
\eq
where $\sigma_{\mu\nu}=(\sigma_\mu+\sigma_\nu)/2$ and $\sigma_\mu$ is
the diameter of the hard-sphere particles of species $\mu$. 
In Eq. (\ref{vc}) we used periodic boundary conditions just to stress
the fact that we are interpreting the Monte Carlo simulations of Das,
Kim, and Fisher \cite{Das2011}, but of course our theoretical
arguments apply to the continuous system as well. 

The system contains $N_\mu$ particles of species $\mu$. We will denote
by $\qq=(\alpha,\rr)$ the species $\alpha$ and the position $\rr$ of a
particle of this species. The particle
$i$ of species $\mu$ has mass $m_{\mu_i}$, charge $z_{\mu_i} e$ with $e$
the unit of charge and $z_{\mu_i}=0,\pm 1,\pm 2, \ldots$,
position $\rr_{i}$, and momentum $\pp_{i}$. The symbol
$\sum^\prime$ means that one should sum over all particles under the
restriction $i\neq j$ when $\alpha_i=\alpha_j$. Periodic boundary
conditions have been assumed in the definition of the
pair-potential. Each charge in the region $\Omega$ is neutralized by a
uniform background of opposite charge density. On account of the
presence of the neutralizing background the term $\kk=\mathbf{0}$ is
excluded in Eq. (\ref{vc}). The potential energy of
Eq. (\ref{modelU}) is defined up to an additive constant, the Madelung
constant $\sum_iz_{\alpha_i}^2\lim_{r\to 0}[v^c(r)-e^2/r]/2$, which
takes into account the interaction of a particle with its own images,
and which becomes important in a grand-canonical calculation. We will
generally use a Greek index to denote the species label and a Roman
index to denote the particle label.

Moreover we impose the constraint 
\bq \label{constraint}
Q=Ne\sum_{\mu=1}^s x_\mu z_\mu=\mbox{constant},
\eq 
where $N=\sum_\mu N_\mu$ is the total number of particles and
$x_\mu=N_\mu/N$ are the molar fractions of particles of species
$\mu$. We also have that $\rho=N/V$ is the particles density 
and $\rho_\mu=\rho x_\mu$ are the partial densities of the ionic
mixture. The neutralizing background has an uniform charge density
$-e\rho_Z$ with $\rho_Z=\rho\sum_\mu x_\mu z_\mu$. 

The 1:1 equisized charge-symmetric hard sphere electrolyte, the RPM
model, is obtained as the particular case with $s=2$, $x_1=x_2=1/2$,
$\sigma_1=\sigma_2=\sigma$, $z_1=-z_2=1$. So that $Q=0$ and the
neutralizing background vanishes.   
 
The RPM has been carefully studied through several computer
simulations and the critical point of the gas-liquid coexistence has
been given various estimates during the years as summarized in Table
\ref{tab:RPM-cp}. On the coexistence spinodal line the isothermal
compressibility $\chi_T=(\partial\rho/\partial
p)_{\{N_\mu\},T}/\rho\to\infty$, with $p$ the pressure of the mixture.  
On approaching the critical point, the amplitude of density
fluctuations increases and local fluctuations become correlated over
increasingly long distances. Anomalies in the intensity of light
scattered from a fluid near its critical point, particularly the
phenomenon known as critical opalescence, were first studied
theoretically by Ornstein and Zernike as far back as 1914
\cite{Fisher1964}. 

\begin{table}
\caption{Critical point estimates for the RPM model from several
  computer simulation 
  studies. The reduced temperature is $T^*=k_BT\sigma/e^2$, with $k_B$
  Boltzmann constant, and the reduced density is $\rho^*=\rho\sigma^3$.}
\label{tab:RPM-cp}
\begin{ruledtabular}
\begin{tabular}{llll}
Reference & year & $T_c^*$ & $\rho_c^*$  \\
\hline
Valleau \cite{Valleau1991}& 1991 & 0.070 & 0.07 \\
Panagiotopoulos \cite{Panagiotopoulos92}& 1992 & 0.056 & 0.04 \\
Orkoulas \cite{Orkoulas1994}& 1994 & 0.053 & 0.025 \\
Caillol \cite{Caillol1996,*Caillol1997}& 1997 & 0.0488(2) & 0.080(5) \\
Orkoulas \cite{Orkoulas1999}& 1999 & 0.0490(3) & 0.070(5) \\
Yan \cite{Yan1999}& 1999 & 0.0492(3) & 0.062(5) \\
Caillol \cite{Caillol2002}& 2002 & 0.04917(2) & 0.080(5) \\
\end{tabular}
\end{ruledtabular}
\end{table}

\section{The moment sum-rules}
\label{sec:sum-rules}

While the thermodynamic stability of the fluid model ensures the
existence of the correlation functions in the thermodynamic limit,  
\bq \nonumber
\rho^{(n)}(\qq_1,\ldots,\qq_n)&=&\rho_1\cdots\rho_n\,
g^{(n)}_{\alpha_1\ldots\alpha_n}(\rr_1,\ldots,\rr_n)\\
&=&\left\langle\sum^\prime_{i_1,\ldots,i_n}
\delta(\rr_1-\rr_{i_1})\delta_{\alpha_1,\alpha_{i_1}}
\cdots\delta(\rr_n-\rr_{i_n})\delta_{\alpha_n,\alpha_{i_n}}
\right\rangle,~~~n=1,2,\ldots,
\eq
where $\langle\ldots\rangle$ is a thermal average defined for an
infinitely extended system, sum-rules are exact relationships that the
correlation functions must obey and can be derived from the
microscopic constituent equations like for example the Born-Green-Yvon
(BGY) hierarchy \cite{Hansen-McDonald} under appropriate plausible
assumptions.

Sometimes it proves convenient to introduce another set of correlation
functions, namely the Ursell's functions $h^{(n)}$,
\bq
g^{(2)}_{\alpha_1\alpha_2}(\rr_1,\rr_2)&=&h^{(2)}_{\alpha_1\alpha_2}(\rr_1,\rr_2)+1,\\ 
\nonumber
g^{(3)}_{\alpha_1\alpha_2\alpha_3}(\rr_1,\rr_2,\rr_3)&=&
h^{(3)}_{\alpha_1\alpha_2\alpha_3}(\rr_1,\rr_2,\rr_3)+
h^{(2)}_{\alpha_1\alpha_2}(\rr_1,\rr_2)+
h^{(2)}_{\alpha_1\alpha_3}(\rr_1,\rr_3)\\
&&+h^{(2)}_{\alpha_2\alpha_3}(\rr_2,\rr_3)+1,\\ \nonumber
g^{(4)}_{\alpha_1\alpha_2\alpha_3\alpha_4}(\rr_1,\rr_2,\rr_3,\rr_4)&=&
h^{(4)}_{\alpha_1\alpha_2\alpha_3\alpha_4}(\rr_1,\rr_2,\rr_3,\rr_4)+
h^{(3)}_{\alpha_1\alpha_2\alpha_3}(\rr_1,\rr_2,\rr_3)+
h^{(3)}_{\alpha_1\alpha_2\alpha_4}(\rr_1,\rr_2,\rr_4)\\ \nonumber
&&+h^{(3)}_{\alpha_1\alpha_3\alpha_4}(\rr_1,\rr_3,\rr_4)+
h^{(3)}_{\alpha_2\alpha_3\alpha_4}(\rr_2,\rr_3,\rr_4)\\ \nonumber
&&+h^{(2)}_{\alpha_1\alpha_2}(\rr_1,\rr_2)h^{(2)}_{\alpha_3\alpha_4}(\rr_3,\rr_4)+
h^{(2)}_{\alpha_1\alpha_3}(\rr_1,\rr_3)h^{(2)}_{\alpha_2\alpha_4}(\rr_2,\rr_4)\\ \nonumber
&&+h^{(2)}_{\alpha_1\alpha_4}(\rr_1,\rr_4)h^{(2)}_{\alpha_2\alpha_3}(\rr_2,\rr_3)+
h^{(2)}_{\alpha_1\alpha_2}(\rr_1,\rr_2)\\ \nonumber
&&+h^{(2)}_{\alpha_1\alpha_3}(\rr_1,\rr_3)+h^{(2)}_{\alpha_1\alpha_4}(\rr_1,\rr_4)\\
&&+h^{(2)}_{\alpha_2\alpha_3}(\rr_2,\rr_3)+h^{(2)}_{\alpha_2\alpha_4}(\rr_2,\rr_4)+
h^{(2)}_{\alpha_3\alpha_4}(\rr_3,\rr_4)+1,\\ \nonumber
\ldots
\eq
It has been shown by Alastuey and Martin \cite{Alastuey1985} that
among all possible long-range potentials, it is only the Coulomb case
that a decay law of the Ursell correlations faster than any inverse
power is compatible with the structure of equilibrium BGY equations.
We may then assume, at least far away from a critical point, that
these Ursell functions tend to zero faster than any power
$r_{ij}^{-m}$ with integer $m$, if the separation $r_{ij}$ between the
positions $\rr_i$ and $\rr_j$ goes to infinity. This assumption is the
usual {\sl exponential clustering hypothesis} for charged systems. 

Introducing the notation $\int d\qq\ldots=\int d\rr
\sum_{\alpha=1}^s\ldots$ we must have the following normalization
properties for the two sets,
\bq
\lim_{N\to\infty} \frac{1}{N^n}\int d\qq_1\ldots
d\qq_n\,\rho^{(n)}(\qq_1,\ldots,\qq_n)&=&1,\\
\lim_{N\to\infty} \frac{1}{N^n}\int d\qq_1\ldots
d\qq_n\,\rho_1\cdots\rho_n h^{(n)}_{\alpha_1\ldots\alpha_n}(\rr_1,\ldots,\rr_n)&=&0.
\eq

In the following we will drop the superscript on the correlation
functions when not leading to confusion. Note also that
$\rho(\qq)=\left\langle\sum_i 
\delta(\rr-\rr_i)\delta_{\alpha,\alpha_i}\right\rangle=\rho_\alpha$ in
a homogeneous mixture whereas
$h^{(2)}_{\alpha_1\alpha_2}(\rr_1,\rr_2)=h_{\alpha_1\alpha_2}(|\rr_1-\rr_2|)$
in a homogeneous and isotropic mixture.
\subsection{The Ornstein-Zernike approach}
\label{sec:OZ}

The Ornstein-Zernike (OZ) equation in reciprocal-space for a fluid
mixture is given by 
\bq
\hat{h}_{\mu\nu}(k)=\hat{c}_{\mu\nu}(k)+\rho\sum_\lambda x_\lambda
\hat{c}_{\mu\lambda}(k) \hat{h}_{\lambda\nu}(k),
\eq
where $k=|\kk|$, $\hat{h}_{\mu\nu}(k)$ is the Fourier transform of the
partial total correlation functions $h_{\mu\nu}(r)=g_{\mu\nu}(r)-1$ with
$g_{\mu\nu}$ the partial radial distribution functions 
\bq \label{rdf}
g_{\mu\nu}(r)&=&\frac{1}{N\rho x_\mu
  x_\nu}\langle\sum^\prime_{i,j}\delta_{\mu,\alpha_i}\delta_{\nu,\alpha_j} 
\delta(\rr-\rr_{i}-\rr_{j})\rangle, 
\eq
and $\hat{c}_{\mu\nu}(k)$ are the Fourier transform of the partial
direct correlation functions \cite{Hansen-McDonald}.

The partial structure factors are defined as 
\bq
S_{\mu\nu}(k)=x_\mu\delta_{\mu\nu}+\rho x_\mu x_\nu \hat{h}_{\mu\nu}(k).
\eq 

Given a partial function $f_{\mu\nu}$ we can now introduce the
following number-number, number-charge, and charge-charge functions 
\bq
\left\{\begin{array}{l}
f_{NN}=\sum_{\mu,\nu}f_{\mu\nu}\\
f_{NZ}=\sum_{\mu,\nu}z_\mu f_{\mu\nu}\\
f_{ZZ}=\sum_{\mu,\nu}z_\mu z_\nu f_{\mu\nu}
\end{array}\right.
\eq
where in the RPM case $f_{NZ}=0$.

We can moreover introduce the following definitions
\bq
\left\{\begin{array}{l}
\tilde{h}_{\mu\nu}=\sqrt{x_\mu x_\nu}\hat{h}_{\mu\nu}\\
\tilde{c}_{\mu\nu}=\sqrt{x_\mu x_\nu}\hat{c}_{\mu\nu}\\
\tilde{S}_{\mu\nu}=S_{\mu\nu}/\sqrt{x_\mu x_\nu}=
\delta_{\mu\nu}+\rho\tilde{h}_{\mu\nu}
\end{array}\right.
\eq
with which the OZ equation can be written in a simple matrix form
\bq \label{OZ-matrix-0}
\mathbf{\tilde{S}}-\mathbf{I}&=&\rho\mathbf{\tilde{S}}
\mathbf{\tilde{c}},
\eq
where $\mathbf{I}$ is the identity matrix. Eq. (\ref{OZ-matrix-0}) can
also be rewritten as follows
\bq \label{OZ-matrix-1}
\mathbf{\tilde{S}}&=&(\mathbf{I}-\rho\mathbf{\tilde{c}})^{-1}.
\eq

It is natural \cite{Hansen-McDonald} to separate the direct
correlation functions into a short-range and a Coulombic part
\bq
\hat{c}_{\mu\nu}(k)=\hat{c}^{sr}_{\mu\nu}(k)-
\frac{4\pi\beta z_\mu z_\nu e^2}{k^2},
\eq
where $\hat{c}^{sr}_{\mu\nu}(k)$ is a regular function in the $k\to 0$
limit. We then see, after some algebra, that in the small $k$ limit,
it must be $S_{NN}\sim k^0$, $S_{NZ}\sim k^2$, and $S_{ZZ}\sim k^2$.   
Moreover, It is a simple algebraic task, starting from the matrix form
$\mathbf{\tilde{S}}=k^2(k^2\mathbf{I}-\rho k^2\mathbf{\tilde{c}})^{-1}$,
to show that for the RPM case 
\bq \label{S-OZ}
S_{ZZ}(k)=\frac{k^2}{(k_D/\bar{z}_2)^2}+
\left(\frac{\rho}{4}\hat{c}^{sr}_{ZZ}(0)-1\right)\frac{k^4}{(k_D/\bar{z}_2)^4}
+O(k^6),
\eq
where $k_D=\sqrt{4\pi\beta\rho \bar{z}_2^2e^2}$ is the Debye
wave-number with $\bar{z}_2^2=\sum_\mu x_\mu z_\mu^2$. In the RPM
$\bar{z}_2^2=1$. Since
we have $S_{ZZ}(k)=\sum_\mu x_\mu z_\mu^2+\rho\sum_{\mu,\nu}x_\mu x_\nu z_\mu
z_\nu\hat{h}_{\mu\nu}(k)$, using spherical symmetry, from
Eq. (\ref{S-OZ}) follow the following first three charge-charge
moment sum-rules 
\bq \label{moment-0zz}
\rho\sum_{\mu,\nu}x_\mu x_\nu z_\mu z_\nu \int d\rr\,h_{\mu\nu}(r)&=&-\bar{z}_2^2\\
\label{moment-1zz}
\rho\sum_{\mu,\nu}x_\mu x_\nu z_\mu z_\nu \int d\rr\,r^2 h_{\mu\nu}(r)&=&
-\frac{6}{(k_D/\bar{z}_2)^2}\\
\label{moment-2zz}
\rho\sum_{\mu,\nu}x_\mu x_\nu z_\mu z_\nu \int d\rr\,r^4 h_{\mu\nu}(r)&=&
-\frac{120}{(k_D/\bar{z}_2)^4}\left(1-\frac{\rho}{4}\hat{c}^{sr}_{ZZ}(0)\right)
\eq
The first identity, the zeroth-moment sum-rule, is a consequence of the
normalization conditions of the correlation functions (\ref{rdf}) 
\bq \label{normalization}
\rho \sum_\mu x_\mu z_\mu\int d\rr\,h_{\mu\nu}(r)=\sum_\mu z_\mu
\frac{\langle N_\mu N_\nu\rangle-\langle N_\mu\rangle\langle
N_\nu\rangle-\delta_{\mu\nu}\langle N_\mu\rangle}{\langle
N_\nu\rangle}=-z_\nu.
\eq 
and reflects {\sl internal} screening (or bulk elecroneutrality). The
second, the second-moment sum-rule, is commonly known as the
Stillinger-Lovett (SL) condition \cite{Stillinger1968} and reflects
{\sl external} screening. The third is the fourth-moment sum-rule.  

In view of the exponential clustering expected to hold in ionic fluids
away from criticality (see next section) we may assume the following
small $k$ expansions
\bq
S_{NN}(k)/S_{NN}(0)&=&1+\sum_{p\ge
  1}(-)^p\xi_{N,p}^{2p}(T,\rho)k^{2p},\\
S_{ZZ}(k)/\bar{z}_2^2&=&0+\xi_{Z,1}^2k^2-\sum_{p\ge
  2}(-)^p\xi_{Z,p}^{2p}(T,\rho)k^{2p},
\eq
where working in the grand-canonical ensemble \cite{Hansen-McDonald}
$S_{NN}(0)=\chi_T/\chi_T^0$ with $\chi_T^0=\beta/\rho$ the isothermal
compressibility of the ideal gas. 

Das, Kim, and Fisher \cite{Das2011} has calculated through
grand-canonical Monte Carlo simulations the second $S_2$ and fourth
$S_4$ moments: $S_{ZZ}(k)/\bar{z}_2^2=0+S_2k^2-S_4k^4+\ldots$ for the
RPM, and found a deviation of about 16\% on the SL condition,
$S_2=1/k_D^2$, at criticality. Moreover $S_4$ appears to diverge to
$+\infty$ upon approaching the RPM critical point. At criticality,
density correlations are long ranged and \cite{Fisher1964}
$S_{NN}(k)\sim 1/k^{2-\eta}$ for $k\to 0$ with $0<\eta<1$ the anomalous
critical-point decay exponent \cite{Fisher1998} (equal to zero in the
Ornstein-Zernike theory). \cite{note2} Equivalently, in real-space, in
three dimensions, $\sum_{\mu,\nu}x_\mu x_\nu h_{\mu\nu}(r)\sim 1/r^{1+\eta}$
for $r\to\infty$. Then according to Proposition 1 of
Ref. \onlinecite{Martin1983} we cannot say anything 
about the SL sum-rule; the fact that the SL sum rule is found to fail
means that the density correlations must decay as $1/r^5$ or
slower. Evidently the development of {\sl clustering or association}
amongst the particles of the mixture upon approaching the critical
point inhibits the external screening. Or in other words, the
diverging density fluctuations that characterize criticality destroy
perfect screening at $(T_c,\rho_c)$.

\subsection{The Born-Green-Yvon approach 
\cite{Suttorp1987,Wonderen1987,Suttorp2008}}
\label{sec:BGY}

Suttorp and van Wonderen \cite{Suttorp1987} study a thermodynamically 
stable ionic mixture with pointwise mobile charges all of the same
sign ($z_\mu\le 0$ for all $\mu$) with the pair-potential of
Eq. (\ref{pp}) without the short-range term $v^{sr}$. Starting from
the Born-Green-Yvon hierarchy \cite{Hansen-McDonald} and using the
hypothesis of exponential clustering of the Ursell's functions they
are able to show that independently of the statistical ensemble used
to describe the ionic liquid the internal screening and SL conditions
(\ref{moment-0zz})-(\ref{moment-1zz}) hold. In
order to make progress for subsequent relationships one has to specify
the ensemble. In a grand-canonical ensemble with the constraint
(\ref{constraint}) the independent variables are $\beta$, $V$,
the $s-1$ chemical potentials, and $q=Q/V$. They are able to
prove the following additional sum-rules for the partial pair Ursell's
functions 
\bq \label{moment-0}
\rho\sum_{\mu,\nu}x_\mu x_\nu \int d\rr\,h_{\mu\nu}(r)&=&
\frac{2}{3}\frac{\beta}{\rho}\frac{\partial\rho}{\partial\beta}-
2\frac{q}{\rho}\frac{\partial\rho}{\partial q}+1,\\ 
\label{moment-2z}
\rho\sum_{\mu,\nu}x_\mu x_\nu z_\mu \int d\rr\,r^2 h_{\mu\nu}(r)&=&
-\frac{6}{(k_D/\bar{z}_2)^2} e
\frac{\partial\rho}{\partial q},\\  
\label{moment-4zz}
\rho\sum_{\mu,\nu}x_\mu x_\nu z_\mu z_\nu \int d\rr\,r^4 h_{\mu\nu}(r)&=&
-\frac{120}{(k_D/\bar{z}_2)^4}\frac{e^2\beta\rho}{q}
\frac{\partial p}{\partial q},
\eq
where $p$ is the pressure and in the partial derivatives all others
independent variables are kept constant. For example, we see that from 
Eq. (\ref{moment-0}) follows
\bq
S_{NN}(0)=\frac{\chi_T}{\chi_T^0}=
\frac{2}{3}\frac{\beta}{\rho}\frac{\partial\rho}{\partial\beta}+ 
2\left(1-\frac{q}{\rho}\frac{\partial\rho}{\partial q}\right).
\eq 

For an ionic mixture with positive and negative mobile charges, made
thermodynamically stable by the addition of the short-range
pair-potential $v^{sr}$, the zeroth-moment of Eqs. (\ref{moment-0})
and (\ref{moment-0zz}) clearly continue to hold as 
well as the second-moment SL sum-rule of
Eqs. (\ref{moment-2zz})-(\ref{moment-2z}) as it is shown in
Ref. \onlinecite{Martin1983}. Note that in order to derive 
the SL sum-rule a weaker condition than the exponential clustering
hypothesis is actually needed as shown in
Ref. \onlinecite{Martin1983}. That is, one just needs to require a
certain short-range behavior of the Ursell functions. For the
fourth-moment condition of 
Eq. (\ref{moment-4zz}) we also expect there to be no effect due to the
short-range regularization as shown in
Ref. \onlinecite{Vieillefosse1985,Vieillefosse1988,Alastuey2016} and
in Appendix \ref{app:2}. So we can say that the Suttorp and van   
Wonderen sum-rules hold generally for the more general ionic liquid
model of a mixture with positive and negative mobile charges
opportunely regularized. 

On the other hand from the work of Santos and Piasecki
\cite{Santos2015} follows that the Ursell functions of any order
have a long-range behavior on a critical point, thus violating
the exponential clustering hypothesis necessary to prove the Suttorp
and van Wonderen sum rules. In this sense the numerical result found
by Fisher {\sl et al.} of the violation of the second and fourth
moment of the charge-charge structure factor of the Restricted
Primitive Model at criticality, is not in contraddiction with the
result of Suttorp and van Wonderen. But is instead telling us
something that goes beyond the analysis of the sum-rules based on the
exponential clustering hypothesis. 

Note that we can write the partial derivative on the right hand side of
Eq. (\ref{moment-2z}) as follows
\bq \nonumber
\frac{\partial\rho}{\partial q}&=&
\frac{\partial(\rho,\mu_1,T,V)}{\partial(q,\mu_1,T,V)}\\
\nonumber
&=&\frac{\partial(\rho,\mu_1,T,V)}{\partial(N_1,N_2,T,V)}
\frac{\partial(N_1,N_2,T,V)}{\partial(q,\mu_1,T,V)}\\
&=&\frac{1}{V}\left[\left(\frac{\partial\mu_1}{\partial N_2}\right)_{N_1}
-\left(\frac{\partial\mu_1}{\partial N_1}\right)_{N_2}\right]_{T,V}
\left[\left(\frac{\partial N_1}{\partial q}\right)_{\mu_1}
\left(\frac{\partial N_2}{\partial\mu_1}\right)_q-
\left(\frac{\partial N_1}{\partial\mu_1}\right)_q
\left(\frac{\partial N_2}{\partial q}\right)_{\mu_1}\right]_{T,V}.
\eq
So that for the symmetric RPM where $\mu_1=\mu_2$, using the
$1\leftrightarrow 2$ 
symmerty, we find $\partial\rho/\partial q=0$, since the first
Jacobian vanishes. Whereas, for a one component system, where
$q=e\rho$, we find $\partial\rho/\partial q=1/e$.     

From the analysis of Suttorp and van Wonderen we also deduce that
\bq \label{S4}
\bar{z}_2^2S_4=\left(\frac{\bar{z}_2}{k_D}\right)^4
\frac{e^2\beta\rho}{q}\frac{\partial p}{\partial q}=
-\left(\frac{\bar{z}_2}{k_D}\right)^4
e^2\beta\rho\frac{\partial^2\tilde{p}}{\partial q^2},
\eq
where $\tilde{p}=p-q\tilde{\mu}_q$ with
$\tilde{\mu}_q=-\partial\tilde{p}/\partial q$ the Lagrange
multiplier which takes into account of the constraint
(\ref{constraint}). The RPM results of Das, Kim, and Fisher
\cite{Das2011} show how $(k_D/\bar{z}_2)^4\bar{z}_2^2S_4\to 0$ for
$\rho\to 0$ (their Fig. 3). This is easily explained observing that as
$\rho\to 0$ we must have $\beta p\to \rho$ so that from Eq. (\ref{S4})
follows 
\bq
(k_D/\bar{z}_2)^4\bar{z}_2^2S_4\to e^2\frac{\partial\rho^2}{\partial
q^2}=0.
\eq
This result also implies that, in view of Eq. (\ref{moment-2zz}),
$\rho\hat{c}_{ZZ}^{sr}(0)\to 4$.

\begin{figure}[htbp]
\begin{center}
\includegraphics[width=8cm]{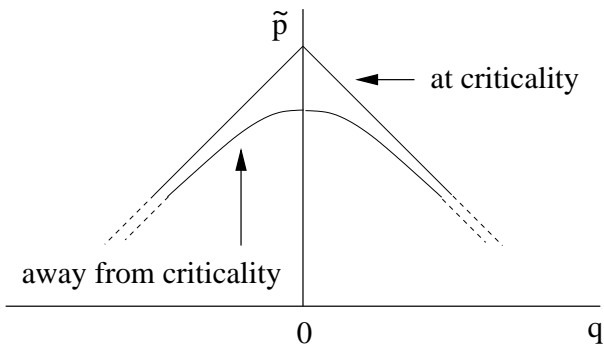}
\end{center}  
\caption{Sketch of $\tilde{p}(q)$ near $q=0$ upon approaching criticality.} 
\label{fig:sketch-p}
\end{figure}

Moreover from Das, Kim, and Fisher \cite{Das2011} Fig. 4, follows that
in the RPM we must have  
\bq
\lim_{q\to 0}\frac{\partial^2 \tilde{p}}{\partial q^2}=-\infty
\eq
when one approaches the critical point. Notice that by
charge symmetry we must have that both $p$ and $\tilde{p}$ are even
functions of $q$. So a sketch of $\tilde{p}(q)$ near $q=0$ must look
as in Fig. \ref{fig:sketch-p}. The figure aims to give a very
qualitative sketch of $\tilde{p}(q)$ only in a very narrow
neighbourhood of $q = 0$. 
Away from criticality we must have $\partial p/\partial q|_{q=0}=0$
and $S_4$ is finite. But near criticality $\partial p/\partial q|_{q=0}>
0$ and $S_4$ diverges. This means that near criticality there is a non
negligible variation of the pressure of the fluid upon switching on a
charge asymmetry ($q\neq 0$) keeping overall neutrality with the
neutralizing background. So notwithstanding the fact that the
exponential clustering hypothesis breaks down near criticality the
results of Das, Kim, and Fisher \cite{Das2011} do not tell us anything
about the failure of the fourth-moment sum-rule. 
On the other hand their Figs. 1 and 2 indicate the failure of the SL
condition upon approaching the critical point, as already observed in
the previous section.

\section{Conclusions}
\label{sec:conclusions}

We studied a general ionic mixture with particles of different mass,
diameter, and charge immersed in a neutralizing background so that
the mixture is globally neutral. When we allow for the presence of
mobile charges of opposite sign we need to add either a soft- or a
hard-core regularization to the pair-potential in order to make the
mixture thermodynamically stable.  

We derived a series of sum-rules on the first three moments of the
charge-charge correlation functions starting from
the Ornstein-Zernike theory \cite{Hansen-McDonald}. Then we showed
that the sum-rules derived by Suttorp and van Wonderen 
\cite{Suttorp1987} for an ionic mixture made of particles all of the
same sign immersed in a neutralizing background remain valid if one
allows the particles to carry charges of opposite sign and adds a
soft or a hard-core repulsion in order to ensure thermodynamic
stability. In particular they remain valid for the symmetric RPM case
when the neutralizing background vanishes. Suttorp and van Wonderen
derivation relies on the assumption of the exponential clustering in
the mixture \cite{Martin1988}.

We interpreted recent results of Das, Kim, and Fisher \cite{Das2011}
reporting the failure of the charge-charge second-moment sum-rules for
the RPM of a ionic liquid at criticality and the
divergence of the charge-charge fourth-moment at criticality. 
In particular the divergence of the fourth moment $S_4$ at the
critical point of the RPM seems to still be in agreement with the
fourth-moment sum-rule (even if the exponential clustering of the
Ursell's function breaks down there as shown in
Ref. \cite{Santos2015}) if one assumes that at 
criticality there is a non negligible variation of the pressure of the 
fluid upon switching on a charge asymmetry ($q\neq 0$) keeping overall
neutrality with the neutralizing background. The observed violation of
the second-moment sum-rule on the other hand seems to indicate that at
criticality the {\sl clustering phenomenon} occurring 
in the ionic mixture is responsible for the break down of the
external screening and the system behaves as an insulator
\cite{Martin1988}. At criticality we do not have anymore an
exponential or short-range clustering but a long-range clustering as
shown by the results of Ref. \cite{Santos2015}.

Our results could be helpfull to a better understanding of
Refs. \cite{Fantoni13e,Fantoni13f} and Refs.
\cite{Fantoni12b,Fantoni03a,Fantoni08c}.  

\appendix

\section{Invariance in form of the moment sum-rules under the addition
  of a hard-core}
\label{app:2}

Let us call PWE the point-wise particle electrolyte considered by
Suttorp and van Wonderen \cite{Suttorp1987} and HSE the hard-sphere
electrolyte obtained by our model of
Eqs. (\ref{model})-(\ref{modelhc}). The configurations space of PWE is  
$\Omega^N$ whereas the one of HSE is
${\cal O}_N=\{\RR\equiv(\rr_1,\ldots,\rr_N)\in\Omega^N|~~~\forall
i,j\neq i
~~~|\rr_i-\rr_j|>\sigma_{\alpha_i\alpha_j}\}\subset
\Omega^N$. In particular it is well known from electrostatics
that HSE is equivalent to the PWE restricted to the configuration
space ${\cal O}_N$. We then conclude that the sum-rules of
Eqs. (\ref{moment-2z}) and (\ref{moment-4zz}) must hold also for the
HSE. In any case the thermodynamic quantities on both sides of the
sum-rule 
will remain unchanged after the restriction. Infact, calling the
complementary set ${\cal O}_N^c=\Omega^N-{\cal
  O}_N=\{\RR\equiv(\rr_1,\ldots,\rr_N)\in\Omega^N|~~~\exists i,j\neq i
~~~|\rr_i-\rr_j|\leq\sigma_{\alpha_i\alpha_j}\}$ we have
for a generic thermal average of an everywhere finite physical
observable 
\bq \nonumber
\langle\ldots\rangle_\text{PWE}&=&
\frac{\int_{\Omega^N}\ldots e^{-\beta U}d\RR}
{\int_{\Omega^N}e^{-\beta U}d\RR}=
\frac{\int_{{\cal O}_N}\ldots e^{-\beta U}d\RR+
\int_{{\cal O}_N^c}\ldots e^{-\beta U}d\RR}
{\int_{{\cal O}_N}e^{-\beta U}d\RR+
\int_{{\cal O}_N^c}e^{-\beta U}d\RR}\\ \nonumber
&=&\frac{\int_{{\cal O}_N}\ldots e^{-\beta U}d\RR\left(1+
\int_{{\cal O}_N^c}\ldots e^{-\beta U}d\RR/
\int_{{\cal O}_N}\ldots e^{-\beta U}d\RR\right)}
{\int_{{\cal O}_N}e^{-\beta U}d\RR\left(1+
\int_{{\cal O}_N^c}e^{-\beta U}d\RR/
\int_{{\cal O}_N}e^{-\beta U}d\RR\right)}\\
&\to&\frac{\int_{{\cal O}_N}\ldots e^{-\beta U}d\RR}
{\int_{{\cal O}_N}e^{-\beta U}d\RR}=
\langle\ldots\rangle_\text{HSE},
\eq
in the thermodynamic limit $\Omega\to\mathbbm{R}^{3}$ and $N=\rho
V$. Since the measure of ${\cal O}_N$ is an infinite of higher order
than the measure of ${\cal O}_N^c$.
This does not mean of course that the Ursell functions themselves will
be equal for the PWE and the HSE and infact they will be different
generally. 

This argument suggests that Suttorp and van Wonderen
analysis \cite{Suttorp1987} continues to hold also for an ionic
mixture with mobile charges of opposite sign opportunely
regularized. This has recently been proved semi-heuristically by
Alastuey and Fantoni \cite{Alastuey2016} for the fourth moment of the 
charge-charge structure factor of such an ionic mixture.

\begin{acknowledgments}
We are grateful to Michael Ellis Fisher and Angel Alastuey for
correspondence and helpful comments. 
\end{acknowledgments}

%

\end{document}